\documentstyle[12pt]{article}

\begin{document}
\begin{center}
{\Large{ \bf Damped Oscillator with $\delta$-kicked frequency in probability representation of quantum mechanics}}
\end{center}
\medskip
{\bf
Vladimir N. Chernega,$^1$ Olga V. Man'ko,$^{1,2}$ }\\
\medskip
{\it $^1$Lebedev Physical Institute, Russian Academy of Sciences\\
Leninskii Prospect 53, Moscow 119991, Russia\\
\smallskip
$^2$Bauman Moscow State Technical University\\
The 2nd Baumanskaya Street 5, Moscow 105005, Russia}\\
\smallskip
$^*$Corresponding author e-mail:~~~omanko\,@\,sci.lebedev.ru

\begin{abstract}\noindent
We obtain the tomogram of squeezed correlated states of a quantum
parametric damped oscillator in an explicit form. We study the
damping within the framework of the Caldirola--Kanai model and chose
the parametric excitation in the form of a very short pulse
simulated by a $\delta$-kick of frequency; the squeezing phenomenon
is reviewed for both models. The cases of strong and weak damping
are investigated.
\end{abstract}
\medskip


\section{Introduction}  
In quantum mechanics, the quantum system state is described by the
wave function (pure state)~\cite{Schrod} or the density matrix
(mixed state)~\cite{Landau,Neumann}; see references in~\cite{dod89}.
In the probability representation of quantum
mechanics~\cite{Mancini,Mancini1,OVVILasRes}, the quantum states are
associated with tomograms. The quantum tomogram is the probability
distribution function of the position $X$ measured in an ensemble of
reference frames. Due to this circumstances, the quantum tomogram
depends on additional parameters $\mu$ and $\nu$ determining the
reference frame in the classical phase space.

The density matrix and wave function of the quantum state can be
reconstructed from the tomogram. The symplectic tomography method
was discussed for measuring quantum states in~\cite{Mancini}. The
method uses the Fourier transform of the marginal distribution for a
measurable squeezed and rotated quadrature (symplectic tomogram)
instead of the Radon transform~\cite{Radon}, which is used in
optical tomography to reconstruct the Wigner function~\cite{Wigner}
from the optical tomogram~\cite{BerBer,VogRis}.

The tomograms determine completely the quantum state and measuring
this distribution implies reconstructing the quantum state.
In~\cite{Raymer}, the scheme of optical tomography was used for
reconstruction of the Wigner function of the electromagnetic field
state. Experiments with reproducible measurements of squeezed vacuum
state of light generated by an optical parametric oscillator were
performed in~\cite{Mlynek}.

The problem of the time-dependent harmonic oscillator is interesting
due to various applications of this model in different areas of
physics. For example, it describes the motion of a single particle
in a Paul trap~\cite{Paul,Brown}. In~\cite{mal73,dod75}, it was
shown that the solutions for systems with quadratic Hamiltonians are
expressed in terms of the classical trajectory of the system.
In~\cite{dod92}, the correlated squeezed states of a quantum
oscillator with $\delta$-kicked frequencies were found, and the
possibility of obtaining the squeezing phenomenon due to the
parametric excitation in the form of $\delta$-kicks of frequency was
shown.

In~\cite{2}, it was shown that any nonadiabatic change in the
frequency of a harmonic oscillator leads to squeezing. The
appearance of squeezing phenomenon due to different types of a
sudden change in frequency was considered in~\cite{6,6b,6c}.
Repeated jumps between two frequencies were shown to generate fast
increasing squeezing if the time intervals between the jumps were
chosen to have special values~\cite{7}. The squeezing properties of
a harmonic oscillator with a series of sudden changes between two
frequencies were calculated in~\cite{kiss}, where infinite,
quasiperiodic, and irregular behaviors of the squeezing were
investigated.

In this paper, we consider the damping within the framework of the
Caldirola--Kanai model~\cite{cal41,kan48}. This model is a partial
case of the multidimensional system described by nonstationary
Hamiltonian, which is a general quadratic form in the position and
momentum operators considered in~\cite{mal73,dod75}. The Gaussian
wave packets in the case of weak damping were investigated
in~\cite{hasse75,caldirola82}. In~\cite{Akpan}, the propagator, wave
function, and uncertainty relation for a time-dependent damped
harmonic oscillator were evaluated, and the classical solution of
the quantum systems was analyzed. In~\cite{Moya-Cessa}, the time
dependent quantum harmonic oscillator subject to a sudden change in
its mass was considered. Squeezing and resonance in a generalized
Caldirola--Kanai-type quantum parametric oscillator was investigated
in~\cite{Buyukask}. In~\cite{Kim}, it was shown that the ground
state of well-known pseudo-stationary states for the
Caldirola--Kanai Hamiltonian is a generalized minimum uncertainty
state, and their dependences on the damping factor and frequency
were found. In~\cite{Choi}, it was shown that a generalization of
the Caldirola--Kanai Hamiltonian, which describes dissipative
systems, can be fulfilled by replacing the standard exponential
function with the $q$-exponential one. The different problems of
quantum oscillator with a time-dependent frequency were solved in
[34--38].    

The aim of this work is to consider the parametric excitation of a
quantum damped oscillator in the probability representation of
quantum mechanics and obtain an explicit form of the tomographic
probability distribution of quadrature in squeezed correlated states
of a quantum damped parametric oscillator, as well as to review the
influence of damping on the squeezing phenomenon for the kicked
oscillator.

This paper is organized as follows.

In Sec.~2, we review the symplectic tomography scheme on the example
of a quantum parametric damped oscillator and obtain symplectic and
optical tomograms of squeezed correlated states. In Sec.~3,
following~\cite{PRChina} we review the appearance of squeezing
phenomenon due to the parametric excitation, which is chosen in the
form of very short pulse simulated by a $\delta$-kick of frequency.
We consider the cases of weak and strong damping and the case of a
free particle. The parametric excitation under consideration is
interesting due to the possibility of finding the classical
trajectory in an explicit form. In Sec.~4, we conclude by listing
the main results obtained in this study.

\section{Tomography of a Quantum Damped Parametric Oscillator}  
It was shown~\cite{Mancini,Mancini1} that for the  generic linear
combination of quadratures, which is a measurable observable (we use
dimensionless variables and put $\hbar=1$),
\begin{equation}\label{X}
\widehat X = \mu \hat q+\nu\hat p\,,
\end{equation}
where $\hat q$ and $\hat p$ are operators of the position and
momentum, respectively. The tomogram $w(X,\mu,\nu)$ (normalized with
respect to the variable $X$), being dependent on the two extra real
parameters $\mu$ and $\nu$, is related to the state of the quantum
system expressed in terms of the density operator $\hat\rho$ as
follows:
\begin{equation}\label{w}
w(X,\mu,\nu) = \mbox{Tr}\left[\hat \rho\delta(X\hat 1-\mu\hat q-\nu\hat p)
\right].
\end{equation}
The physical meaning of the parameters $\mu$ and $\nu$ is that they
describe an ensemble of rotated and scaled reference frames, in
which the position $X$ is measured. For $\mu =\cos\varphi$ and
$\nu=\sin\varphi$, the tomogram~(\ref{w}) is the distribution of
homodyne output variable used in optical
tomography~\cite{BerBer,VogRis}. Formula~(\ref{w}) can be inverted,
and the density operator of the state can be expressed in terms of
the tomogram~\cite{Marmo},
\begin{equation}\label{W}
\hat\rho = \frac{1}{2\pi}\int w(X,\mu,\nu)e^{i(X-\mu\hat q-\nu\hat p)}
  \,dX\,d\mu\,d\nu.
\end{equation}
First, we discuss the tomogram for squeezed and correlated state of
a damped parametric oscillator; it has the Gaussian form,
\begin{equation} \label{freesolution}
w_\alpha(X,\mu,\nu,t) = \frac{1}{\sqrt{2\pi\sigma_X(t)}}\,
  \exp\left\{-\frac{(X-\bar{X})^2}{2\sigma_X(t)}\right\},
\end{equation}
in which, in view of (\ref{X}), one can express the mean value of
observable as follows:
\begin{equation} \label{insert7}
\bar{X} = \mu\langle q\rangle +\nu\langle p\rangle \,,
\end{equation}
where $\langle p\rangle$ and  $\langle q\rangle$ are the quadrature
means in the state~(\ref{freesolution}), and $\alpha$ is a complex
number.

The dispersion of the observable $X$ is
\begin{equation}\label{insert6}
\sigma_X(t)=\mu ^2\sigma_{q^2}+\nu ^2\sigma_{p^2}
+2\mu \nu \sigma_{pq}\,,
\end{equation}
where the parameters $\sigma_{q^2}\,,\,\sigma_{p^2}\,,$ and
$\sigma_{pq}$ are dispersions and covariance of quadratures in the
state~(\ref{freesolution}). The tomogram~(\ref{freesolution}) is the
probability distribution function; it is normalized, and completely
determines the squeezed correlated quantum state of a damped
parametric oscillator.

Let us consider the quantum damped parametric oscillator within the
framework of the Caldirola--Kanai model~\cite{cal41,kan48}. The
Hamiltonian of the system reads
\begin{equation}\label{eq.1}
\widehat H = \frac {1}{2} m e^{2\gamma t} \omega^{2}(t) {\widehat q}^{2} +
\frac{1}{2m} e^{-2\gamma t} {\widehat p}^{2},
\end{equation}
where $m$ is the mass of oscillator, $\gamma$ is damping
coefficient, $\widehat q$ is the position operator, $\widehat p$ is
the momentum operator, and $\omega(t)$ is the time-dependent
frequency of oscillator. For simplicity, we assume put $m=1$ and
$\hbar=1$.

The classical equations of motion for the classical position $q$ and
momentum $p$ corresponding to the Caldirola--Kanai model are
\begin{equation}\label{eq.2}
\dot q= p e^{-2\gamma t},\qquad \dot p = - \omega^{2}(t) e^{2\gamma
t}q,\qquad
\ddot q + 2\gamma \dot q + \omega^{2}(t) q = 0.
\end{equation}
We consider the function $\varepsilon(t)$, which is the solution of
the equation of motion
\begin{equation}\label{eq.4}
\ddot \varepsilon(t) + 2\gamma \dot \varepsilon(t) + \omega^{2}(t)
\varepsilon(t) =0,\qquad
\Omega^{2}(0) = \omega^{2}(0) - \gamma^{2},
\end{equation}
with the initial conditions $\varepsilon(0) = 1,~ \dot
\varepsilon(0) =
\dot\imath \Omega(0)$, and it satisfies the constraint
\begin{equation}\label{eq.5}
e^{2\gamma t} \left( \dot\varepsilon \varepsilon^{\ast} - \dot\varepsilon^{\ast}
\right)= 2 \,\dot\imath \, \Omega(0).
\end{equation}
The quantum dispersion of the position in state~(\ref{freesolution})
is expressed through the classical trajectory $\varepsilon(t)$
(\ref{eq.4})
\begin{equation}\label{eq.7}
\sigma_{q^2} ={\mid \varepsilon \mid^2}/{2}.
\end{equation}
For the quantum dispersion of the momentum in the state
(\ref{freesolution}), one can obtain the following expression:
\begin{equation}\label{eq.8}
\sigma_{p^2} =
\frac{ e^{4 \gamma t} \mid \dot\varepsilon \mid^2}{2 \, \Omega^2(0)}.
\end{equation}
The covariance of the position and momentum in the state
(\ref{freesolution}) is
\begin{equation}
\sigma_{q p}=\frac{1}{2}\sqrt{\frac{e^{4\gamma t}|\varepsilon\dot\epsilon|^2}{\Omega^2(0)}-1}.
\end{equation}
The mean values of quadratures in the state (\ref{freesolution}) are
\begin{equation}\label{insert5a}
\langle p\rangle =\frac {e^{2\gamma t}}{\sqrt 2\Omega(0)}
\left(\alpha \dot \varepsilon ^*-\alpha ^*\dot \varepsilon \right)\,,\qquad
\langle q\rangle =\frac {1}{\sqrt 2}\left(\alpha \varepsilon ^*
+\alpha ^*\varepsilon \right).
\end{equation}
The mean value of observable $X$~(\ref{insert7}) in squeezed
correlated states~(\ref{freesolution}) is expressed through the
classical trajectory as follows:
\begin{equation}\label{meanX}
\bar X=\frac{1}{\sqrt 2}
\left[\mu\left(\alpha\varepsilon^\ast+\alpha^\ast\varepsilon\right)+
\nu \frac{e^{2\gamma t}}{\Omega(0)} \left(\alpha\dot\varepsilon^\ast-\alpha^\ast\dot\varepsilon\right)\right].
\end{equation}
For dispersion of observable $X$ (\ref{insert6}) in squeezed
correlated states (\ref{freesolution}), one has
\begin{equation}\label{disobs}
\sigma_X=\frac{1}{2}\left(\mu^2|\varepsilon|^2+\frac{\nu^2}{\Omega^2(0)}
e^{4\gamma t}|\dot\varepsilon|^2+\nu\mu\sqrt{\frac{e^{4\gamma t}|\epsilon\dot\epsilon|^2}{\Omega^2(0)}-1}\right).
\end{equation}
The explicit expression for the symplectic tomogram, which
determines the squeezed correlated states of a quantum damped
parametric oscillator, reads
\begin{eqnarray}
&&w_\alpha\left(X,\,\mu,\,\nu \right
)=\left[\pi\left(\mu^2|\varepsilon|^2
+\frac{\nu^2}{\Omega^2(0)}e^{4\gamma t}|\dot\varepsilon|^2
+\mu\nu\sqrt{\frac{e^{4\gamma
t}|\epsilon\dot\epsilon|^2}{\Omega^2(0)}-1}
\right)\right]^{-{1}/{2}}\nonumber\\
&&\times\exp\left[-\frac{\left(X-
\mu\left\{\left(\alpha\varepsilon^\ast+\alpha^\ast\varepsilon\right)/\sqrt2\right\}-
\left\{\nu e^{2\gamma t}
\left(\alpha\dot\varepsilon^\ast-\alpha^\ast\dot\varepsilon\right)/\sqrt2\Omega(0)\right\}
\right)^2}{\mu^2|\varepsilon|^2+[\nu^2e^{4\gamma
t}|\dot\varepsilon|^2/\Omega^2(0)]
+\nu\mu\sqrt{e^{4\gamma
t}|\epsilon\dot\epsilon|^2/\Omega^2(0)-1}}\right].\label{symplectictom}
\end{eqnarray}
The optical tomogram of squeezed correlated states of quantum damped
parametric oscillator has the Gaussian form
\begin{eqnarray}
&&w_\alpha^{\rm opt}\left(X,\,\theta \right
)=\left[\pi\left(\cos^2\theta|\varepsilon|^2
+\frac{\sin^2\theta}{\Omega^2(0)}e^{4\gamma t}|\dot\varepsilon|^2
+\sin\theta\cos\theta\sqrt{\frac{e^{4\gamma
t}|\epsilon\dot\epsilon|^2}{\Omega^2(0)}
-1}\right)\right]^{-{1}/{2}}\nonumber\\
&&\times\exp\left[-\frac{\left\{X-
\cos\theta\left[\left(\alpha\varepsilon^\ast+\alpha^\ast\varepsilon\right)/\sqrt2\right]-
\sin\theta e^{2\gamma t}\left[
\left(\alpha\dot\varepsilon^\ast-\alpha^\ast\dot\varepsilon\right)/\sqrt2\Omega\,(0)\right]
\right\}^2}{\cos^2\theta|\varepsilon|^2+[\sin^2\theta e^{4\gamma
t}|\dot\varepsilon|^2/\Omega^2(0)]
+\sin\theta\cos\theta\sqrt{e^{4\gamma
t}|\epsilon\dot\epsilon|^2/\Omega^2(0)-1}}\right].\label{opttom}
\end{eqnarray}
The optical tomogram (\ref{opttom}) satisfies the entropic
inequality
\begin{equation}\label{inequality}
-\int\left[w_\alpha^{\rm opt}\left(X,\,\theta \right )\ln
w_\alpha^{\rm opt}\left(X,\,\theta \right )+w_\alpha^{\rm
opt}\left(X,\,\theta+\pi/2 \right )
\ln w_\alpha^{\rm opt}\left(X,\,\theta+\pi/2 \right )\right] d\,X\geq\ln(\pi
e).
\end{equation}
This inequality can be checked experimentally in the optical
tomography measurement scheme.

The squeezing coefficient in the states determined by
tomograms~(\ref{symplectictom}) and (\ref{opttom}) reads
\begin{equation}\label{eq.9}
k= {\sigma_{q^2}(t)}/{\sigma_{q^2}(0)} = \mid \varepsilon \mid^2.
\end{equation}
It is also expressed through the classical trajectory
$\varepsilon(t)$. In the case where $\mid \varepsilon \mid^2$ is
smaller then unity, which means that the dispersion of position at
the some moment of time $t$ is less than that at the initial moment
of time, the squeezing phenomenon appears. Due to this, the
states~(\ref{freesolution}) have the name of squeezed correlated
states, as well as in the case without damping. So one can see that
all physical characteristics of the system are expressed through the
solution of the classical equation of motion $\varepsilon(t)$.  The
only remaining problem is to find an explicit expression for the
function $\varepsilon$. We devote the following sections to finding
the explicit expressions for classical trajectories for different
regimes of damping.

\section{Damped Oscillator with Parametric Excitation in the Form
of a  $\delta$-Kick of Frequency}
We consider a quantum damped oscillator with the time-dependent
frequency, which is varied in a special manner of $\delta$-kick
$$\omega^2(t) = \omega^2_0 - 2\kappa\delta(t), $$
where $\omega_0$ is the constant part of frequency, and $\delta$ is
the Dirac delta-function. We have the following equation for the
function $\epsilon(t)$:
\begin{equation}\label{eq.10}
\ddot\epsilon(t) + 2\gamma \dot\epsilon(t) + \omega_0^2 \epsilon(t)
-2 \kappa\delta(t) = 0.
\end{equation}

\subsection{The Case of Weak Damping}
In this section, we consider the case of weak damping when
$\omega_0>\gamma$.  Before and after the kick of frequency, the
solution for (\ref{eq.10}) is given by
\begin{equation}\label{eq.11}
\epsilon_k(t) = A_k e^{-\gamma t + \dot\imath \Omega t} + B_k e^{-\gamma t
-\dot\imath \Omega t},\qquad k=0,1.
\end {equation}
where $\Omega = (\omega_0^2 -\gamma^2 )^{1/2}.$ Due to the
continuity conditions,
\begin{equation}\label{eq.12}
\epsilon_{0}(0) = \epsilon_1(0),\qquad \dot\epsilon_1(0) -
\dot\epsilon_{0}(0) = 2\kappa \epsilon_{0}(0).
\end{equation}
The coefficients $A_k$ and $B_k$ must satisfy the relations, which
in the matrix form read
\begin{equation}\label{eq.13}
=\left(\begin{array}{c}
  A_1  \\  B_1
  \end{array}\right)
=\left(\begin{array}{cc}
  1-{\dot\imath\kappa}/{\Omega} & -{\dot\imath\kappa}/{\Omega}\\
  {\dot\imath \kappa}/{\Omega} & 1+{\dot\imath\kappa}/{\Omega}
  \end{array}\right)
  \left(\begin{array}{c}
   A_0 \\ B_0
  \end{array}\right).
\end{equation}
If $\epsilon(-0)=1$ and  $\dot\epsilon(-0)=\dot\imath\Omega$ at the
initial time instant, then $A_0=1-{\dot\imath\gamma}/{2\Omega}$ and
$B_0={\dot\imath\gamma}/{2\Omega}$, and the classical trajectory
after the kick can be expressed as
\begin{equation}\label{eq.14}
\epsilon_{1}(t) = \left(1-\frac{\dot\imath(\kappa+\gamma/2)}{\Omega}\right)
\exp(-\gamma t+\dot\imath\Omega t)-
\frac{\dot\imath(\kappa+\gamma/2)}{\Omega}\exp(-\gamma t-\dot\imath\Omega t).
\end{equation}
If before the first $\delta$-kick of the frequency the quantum
oscillator was in the coherent state determined by
formula~(\ref{symplectictom}) with
\begin{equation}\label{do}
\epsilon(t)=e^{-\gamma t}\left[e^{\dot\imath\Omega t}+
({\gamma/\Omega})\sin\Omega t\right],
\end{equation}
which can be considered as the coherent state of a damped oscilator
due to the analogy with undamped case, then the parametric
excitation will transform it into a squeezed correlated state
determined by the tomogram~(\ref{symplectictom}) with the function
$\epsilon(t)$ given by~(\ref{eq.14}). It is not difficult to
calculate the quantum dispersion of position in excited correlated
squeezed state; it reads
\begin{equation}\label{eq.15}
\sigma_{q^2}(t)=\frac{e^{-2\gamma t}}{2\,\Omega}\left[ 1+
\frac{\sin^2\Omega t}{\Omega^2}(2\kappa+\gamma)^2 +
(2\kappa+\gamma)\frac{\sin 2\Omega t}{\Omega} \right].
\end{equation}
From the above expressions, we see that the maximum and minimum of
$\sigma_{q^2}(t)$ and of squeezing coefficient
$k^2(t)=\sigma_{q^2}(t)/\sigma_{q^2}(0)$ depend on the ratios of
force of delta-kick and damping constant to the frequency of
oscillations, while the lower limit of the squeezing coefficient is
\begin{eqnarray}
&&k^2=\left[ 1 +2\frac{(\kappa+\gamma/2)^2}{\Omega^2} -
2\frac{(\kappa+\gamma/2)}{\Omega^2}\sqrt{(\kappa+\gamma/2)^2+\Omega^2}
\right]\nonumber\\
&&\times\exp\left[ \frac{\gamma}{\Omega}
\cos^{-1}\left(\frac{\Omega}{\sqrt{(\kappa+\gamma/2)^2+\Omega^2}}\right)-
\frac{\pi\gamma}{\Omega}(2n-1) \right],\quad n=0,1,\ldots\label{eq.16}
\end{eqnarray}
From the above formulas, one can see that the squeezing phenomenon
can be achieved for all values of the damping coefficient. So
choosing more strong kicks of frequency (increasing the force of the
delta-kick), we can squeeze quantum noise in the oscillator position
even in the case of large (but less than $\omega_0$) damping
coefficient $\gamma$.

\subsection{Strong Damping}
Now we consider a quantum damped oscillator in the regime of strong
damping, where $\gamma>\omega_{0}$. In this case, the solution to
Eq.~(\ref{eq.10}) before and after the kick of frequency is
$\epsilon_{k}=A_{k}e^{(\Omega -\gamma)t} + B_{k}e^{-(\gamma + \Omega
)}t$, with frequency $\Omega = (\gamma^{2} -\omega_{0}^{2} )^{1/2}$.
Using the same procedure as in the previous section, one can obtain
that after the $\delta$-kick of frequency the coefficients
$A_{1},~B_{1}$ are connected with the initial ones in the same way
through the matrix equation,
\begin{equation} \label{eq.17}
\left( \begin{array}{c}
  A_1 \\ B_1
  \end{array}\right) =
  \left(\begin{array}{cc}
  1+ {\kappa/\Omega} & {\kappa/\Omega} \\
  -{\kappa/\Omega} & 1-{\kappa/\Omega}
  \end{array}\right)
  \left(\begin{array}{cc}
  A_0 \cr B_0
  \end{array}\right).
\end{equation}
Taking the initial conditions in the form $\epsilon(0)=1$ and
$\dot\epsilon(0)=\dot\imath\Omega$, one arrives at
$A_0=\displaystyle{1\over 2}(1+\dot\imath+\gamma/\Omega)$ and
$B_0=\displaystyle{1\over2}(1-\dot\imath-\gamma/\Omega)$, and obtain
for the classical trajectory $\epsilon(t)$ after the $\delta$-kick
of frequency the following expression:
\begin{equation}\label{eq.18}
\epsilon(t) =  e^{-\gamma t}\left[\cosh \Omega t +\sinh\Omega t\left(\dot\imath
+\frac{\gamma}{\Omega} +\frac{2\kappa}{\Omega}\right)\right].
\end{equation}
If before the first $\delta$-kick of frequency the quantum
oscillator was in the coherent state determined by
(\ref{symplectictom}) with function $\varepsilon$ determined by
(\ref{do}), then after the $\delta$-kick of frequency it occurs in
squeezed correlated state determined by
tomogram~(\ref{symplectictom}) with function $\epsilon(t)$ given by
(\ref{eq.18}), and the dispersion of position takes the form
\begin{equation}\label{eq.19}
\sigma_{q^2}(t)=\frac{\hbar e^{-2\gamma t}}{2\,\Omega}\left[ \cosh 2\,\Omega t +
\left(\frac{2\kappa+\gamma}{\Omega}\right)^2\frac{\cosh 2\,\Omega t -1}{2} +
\frac{2\kappa+\gamma}{2}\sqrt{\cosh^2 2\,\Omega t-1}\right].
\end{equation}
Due to the property of $\cosh \alpha$ not being less than unity, one
can see that the dispersion (\ref{eq.19}) can never be less than
$\frac{\hbar e^{-2\gamma t}}{2\, \Omega}$, so the squeezing
phenomenon cannot be achieved in the system under study by a
$\delta$-kick of frequency in the regime of strong damping.

\subsection{Parametric Excitation of Free particle Motion}
In this section, we consider the case where the constant part of
frequency is equal to zero but the parametric excitation acts on the
free particle motion. The linear integrals of motion and Gaussian
wave packets for such systems without parametric excitation were
considered in~\cite{hasse75}. In the case $\omega_{0}=0$, the
equation for classical trajectory reads
\begin{equation}\label{eq.20}
\ddot\epsilon(t) + 2\gamma\dot\epsilon(t) - 2\kappa \delta(t)=0.
\end{equation}
Before and after the delta-kick of frequency, the solution for this
equation is given by the expression $\epsilon_{k}=A_{k}
+B_{k}e^{-2\gamma t}.$ Applying the procedure used before and the
continuity conditions, one can obtain the relation between the
coefficients after $\delta$-kick and the initial ones in the form
\begin{equation}\label{eq.21}
\left( \begin{array}{c}
  A_1 \\ B_1
  \end{array}\right) =
  \left(\begin{array}{cc}
  1+{\kappa/\gamma} & {\kappa/\gamma} \\
  -{\kappa/\gamma} & 1-{\kappa/\gamma}
  \end{array}\right)
  \left(\begin{array}{c}
  A_0 \\ B_0
  \end{array}\right).
\end{equation}
Taking into account the coefficients
$A_0=1+\dot\imath/2~B_0=-\dot\imath/2$, which coincide with the
initial conditions considered above, the expression for classical
trajectory after the kick can be obtained as follows:
\begin{equation}\label{eq.22}
\epsilon(t)=1+{\kappa \over\gamma}(1-e^{-2\gamma t})+
{\dot\imath\over 2}(1-e^{-2\gamma t}).
\end{equation}
The excited states are determined by tomogram~(\ref{symplectictom}),
where the function $\varepsilon$ is given by (\ref{eq.22}), and the
squeezing coefficient is equal to
\begin{equation}\label{eq.23}
k^2=1+({\kappa^2/\gamma^2})(1-e^{-2\gamma t})^2+
2\,\frac{\kappa}{\gamma(1-e^{-2\gamma t})}.
\end{equation}
From expression (\ref{eq.23}), we can see that the squeezing
coefficient $k^2$ can never be smaller than unity because the
function $e^{-2\gamma t}$ is always greater than unity since we
investigate the problem in positive moments of time ($t>0$), and the
damping coefficient is a positive number ($\gamma>0$). The squeezing
phenomenon cannot be achieved for free damped particle by a one-kick
of frequency. In the case of zero damping ($\gamma=0$) and in the
limit of free particle ($\omega_0=0$), one $\delta$-kick of
frequency does not produce any squeezing~\cite{mielnik}.

\section{Conclusions}
In the probability representation of quantum mechanics and within
the framework of the Caldirola--Kanai model, we considered the
parametric excitation of a damped oscillator and obtain the
tomograms of squeezed correlated states in an explicit form;
formulas~(\ref{symplectictom}) and (\ref{opttom}). We discussed the
influence of different regimes of damping on the possibility of
appearance of squeezing phenomenon in the system under study. We
mention the possibility of appearance of squeezing phenomenon due to
the parametric excitation chosen in a special form of a
$\delta$-kick of frequency in the case of weak damping (for all
$\gamma$ less than $\omega_0$) by choosing different forces of the
kick. In the regime of strong damping and for free particle motion,
it is impossible to reach the squeezing phenomenon by a
$\delta$-kick of frequency.

It is worth noting that the results of papers~[20,~23,~24,~38]
performed under the supervision of Professor Janszky provided
important contribution in the theory of the phenomena under
discussion.

We were happy to have the opportunity to visit Hungary in connection
with the 5th International Conference on Squeezed States and
Uncertainty Relations (Lake Balaton, Hungary, May 27--31, 1997) and
the Wigner Centennial Conference (Pecs, Hungary, July 8--12, 2002)
organized by Professor Janszky, and we dedicate this article to his
memory.

\end{document}